\def \manuflag {0}
 
\ifnum \manuflag = 0
 \documentclass[pre,twocolumn,showpacs,amsmath,amssymb]{revtex4}
 \usepackage{psfig}	
 \usepackage{epsfig}		
 \usepackage{delarray}
 \usepackage{graphicx}
 \usepackage{dcolumn}
 \usepackage{bm}
 \def \Title{
Universal $1/f$ noise, cross-overs of scaling exponents, 
and chromosome specific patterns of GC content in
DNA sequences of the human genome}

 \newcommand{\SEC}{\section}
 
\else
 \documentclass[pre,onecolumn,showpacs,amsmath,amssymb]{revtex4}
 \usepackage{psfig}	
 \usepackage{epsfig}		
 \usepackage{delarray}
 \usepackage{graphicx}
 \usepackage{dcolumn}
 \usepackage{bm}
 \usepackage{rotating}
 \def \Title{Universal $1/f$ Noise, crossing-Overs of scaling
exponents, and chromosome specific patterns of GC content in
DNA sequences of the human genome}

\fi

\def \Abstract{
Spatial fluctuations of guanine and cytosine base content (GC\%) are
studied by spectral analysis for the complete set of human genomic DNA
sequences. We find that (i) the $1/f^{\alpha}$ decay is universally
observed in the power spectra of all twenty-four chromosomes, and that
(ii) the exponent $\alpha \approx 1$ extends to about $10^7$ bases,
one order of magnitude longer than what has previously been observed.
We further find that (iii) almost all human chromosomes exhibit a
cross-over from $\alpha_1 \approx 1$ ($1/f^{\alpha_1}$) at lower
frequency to $\alpha_2 < 1$ ($1/f^{\alpha_2}$) at higher frequency, 
typically occurring at around 30,000--100,000
bases, while (iv) the cross-over in this frequency range is 
virtually absent in
human chromosome 22. In addition to the universal $1/f^\alpha$ noise in power
spectra, we find (v) several lines of evidence for chromosome-specific
correlation structures, including a 500,000 bases long oscillation in
human chromosome 21. The universal $1/f^\alpha$ spectrum in human 
genome is further substantiated by a resistance to variance reduction 
in guanine and cytosine content when the window size is increased.
}

\ifnum \manuflag = 0
 
\begin{document}
 \title{\Title}
 \author{Wentian Li}
 \email{wli@nslij-genetics.org}
 \affiliation{The Robert S. Boas Center for Genomics and Human Genetics,
	      North Shore LIJ Institute for Medical Research, 350 Community Dr.,
	      Manhasset, NY 10030.}
 \author{Dirk Holste}
 \email{holste@mit.edu}
 \affiliation{Department of Biology, 
	      Massachusetts Institute of Technology, Cambridge, MA 02139.}
 \begin{abstract}
   \Abstract
 \end{abstract}
 \pacs{87.10.+e, 87.14.Gg, 87.15.Cc, 02.50.-r, , 02.50.Tt, 89.75Da, 89.75.Fb, 05.40.-a}
 %% \keywords{Suggested keywords if desired}
 \maketitle
\else
 \begin{document}
 \title{\Title}
 \author{Wentian Li}
 \email{...@...}
 \affiliation{...}
 \author{Dirk Holste}
 \email{holste@mit.edu}
 \affiliation{Department of Biology,
	      Massachusetts Institute of Technology, Cambridge, MA 02139.}
 \begin{abstract}
   \Abstract
 \end{abstract}
 \pacs{87.10.+e, 02.50.-r, 05.40.-a  \hfill {\tt Thu Sep  5 15:36:39 EDT 2002}}
 %% \keywords{Suggested keywords if desired}
\maketitle
\fi

\SEC{Introduction}

By measuring the proportion of a signal's power $S(f)$ falling into
a range of frequency components $f$, a power spectrum of the form
$S(f) \sim  1/f^\alpha$ distinguishes between two prototypes of noise:
white noise ($\alpha = 0$) and Brownian noise ($\alpha = 2$). The
intermittent range, termed ``$1/f$ noise'', can practically be defined
as $1/f^\alpha$ ($0.5 \lesssim \alpha \lesssim 1.5$). $1/f$ noise was experimentally 
observed first in electric current fluctuations of the thermionic 
tube at the beginning of the nineteenth century \cite{johnson}.
Since then, $1/f$ noise has been found repeatedly in many other 
conducting materials \cite{1freview}.  More generally, it has also 
been observed in wide ranges of natural as well as human-related
phenomena, including traffic flow, star light, speech, music 
and human coordination \cite{1freview-other,1fbib}.
For biological sequences, such as DNA, the concept of slow-varying, 
multiple-length variations in the power of frequency components 
can be translated to long-ranging correlations in the spatial 
arrangement of the four bases adenine (A), cytosine (C), guanine 
(G) and thymine (T).  One can categorize chemically A, C,
G, and T as strong (G or C) or weak (A or T) bonding. It has been
shown that fluctuations of the GC base content along a DNA sequence
are typically stronger correlated when compared to other possible
binary classifications \cite{mapping,dirk}. 
Initial studies of $1/f$ noise in DNA sequences were motivated 
by a model of spatial $1/f$ noise of symbolic sequence evolution 
\cite{wli-em}.  Subsequently, empirical $1/f$ spectra were 
indeed observed in non-protein-coding DNA sequences 
\cite{wli-dna}, and their generality in DNA sequences was further 
illustrated in \cite{voss}.

$1/f$ noise has been detected in a variety of different species and
taxonomic classes, including bacteria \cite{bac}, yeasts
\cite{yeast}, insects \cite{fukushima}, and other higher eukaryotic
genomes. Integrating this and several other lines of evidence, a
consensus on $1/f$ noise in DNA sequences has emerged: 
(1) for DNA sequences of the order of $10^6$~bases ($1$~Mb), $1/f^\alpha$ 
spectrum ($\alpha \approx 1$) is consistently observed; (2) for 
isochores, which are DNA sequences of relatively homogeneous base 
concentration at least $300\cdot 10^3$ bases ($300$~kb) long 
\cite{isochore,isochore-clay,cc03}, $1/f^\alpha$ spectrum is also
observed,  but typically shows a smaller exponent $\alpha < 0.7$
\cite{isochore-clay,clay3,isochore-spe}; (3) for DNA sequences of
the order of several kb, the decay of $S(f)$ is non-trivial and may
depend on whether the sequence is protein-coding \cite{wli-dna}. 
The viral DNA sequence of the $\lambda$-phage, e.g., shows a single step in its GC
base concentration and its spectrum is $S(f) \sim 1/f^2$, which is
characteristic of random block sequences \cite{wli-complexity}. 
We note that the universal scaling of $S(f) \sim 1/f^\alpha$ ($\alpha \approx 1$)
across all species discussed in \cite{voss} has apparently been
restricted to a length scale of $1$~kb, by averaging the spectrum over
many $N=2$~kb DNA segments.

With the availability of the first completed version of the DNA
sequence of human genome \cite{lander}, several studies have been able
to demonstrate that the base-base correlation function $\Gamma(d)$
($d$ distance between bases) of several DNA sequences follows a
power-law decay, $\Gamma(d) \sim 1/d^{\gamma}$. For instance, the DNA
sequence of human chromosome 22 shows statistically significant
power-law correlations up to $d=1$~Mb, and
correlations in the DNA sequence of chromosomes 21 are statistically
significant up to several~Mb (with the scaling exponent $\gamma$
changing beyond a few~kb) \cite{dirk,pedro}.
While the DNA sequences of human chromosomes 21 and 22 are about
$34$~Mb long, in order to estimate the limit of the range of $1/f^\alpha$
spectrum, longer sequences are necessary.

After the release of the draft of the human genome sequence in February 
2001, about three years later in 2004, a dozen (out of 24) human chromosomes 
have been completed with a sequence accuracy to following the 
standard of less than one error per 10,000 DNA
bases (99.99\% accuracy) \cite{hg-quality}.  Building upon the release
of updated, high-quality sequence data, in the era of genomics we can
now conduct a systematic analysis of several issues of $1/f$ noise in
the DNA sequences of our own species {\em Homo sapiens}, which 
have been pursued over the last decade in a fragmentary manner.

In this paper, we use the DNA sequences of the complete set of
twenty-two autosomes and two sex chromosomes to address the following 
issues: Is $1/f$ noise
universally present across the entire set of human genome sequences?
Does $1/f$ noise extend to lower frequency ranges in longer DNA
sequences? Is the decay of $S(f)$ characterized by a single exponent
$\alpha$, or does it exhibit cross-overs (multiple scaling exponents)? 
Given the presence of universal variations at multiple scales, 
do these co-exist with variations at chromosome-specific scales?

\begin{figure}
\centerline{\psfig{figure=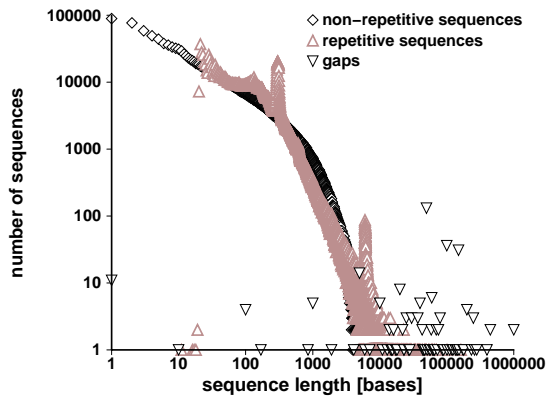,width=68mm,angle=0}}
\caption{\label{fig-dirk1}
Double-logarithmic representation of the human genome-wide length
distribution of interspersed repeat sequences, non-repetitive
sequences, and sequences of unknown base composition (gaps).  The
length distribution of interspersed repeats and non-repetitive sequences
exhibits a power-law-like decay, while that of gap sequences
is scattered across different sequence length. The peaks 
at $\sim 300$ bases and
several kb correspond to Alu and possibly LINE repeats.
}
\end{figure}

\begin{figure}[ht]
\centerline{\psfig{figure=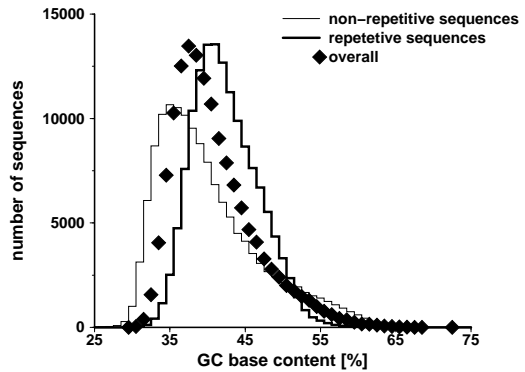,width=68mm,angle=0}}
\caption{\label{fig-dirk2}
Distribution of genome-wide GC content (GC\%) of the human genome
for interspersed repeat sequences, non-repetitive sequences, 
and all (``overall") sequences with sequence segments of $20~{\rm kb}$. 
The mode (peak location) of non-repetitive sequences is at 
$\sim$35\%, while the mode of repetitive sequences shifted 
to a higher GC\% ($\sim$42\%). The fraction 
of non-repetitive sequences with GC\%~$>$~50\% is markedly 
larger as compared to the repetitive sequences.
}
\end{figure}

\SEC{Data and methods}

In this section, we introduce the data for human genome sequences, as well
as the notation and definitions used throughout this study.
Twenty-four chromosomes are assembled in build 34
of the NCBI (human genome hg16 release). Sequence data were downloaded
from the UCSC human genome repository (available at {\sf http://genome.ucsc.edu/}).
Unsequenced bases are kept to preserve spacing between 
bases. Human chromosomes (Chr) 13, 14, 15, 21, and 22 contain large 
amount of unsequenced bases in the left end of their DNA sequences, 
consisting of about 15\%, 17\%, 18\%, 21\%, and 29\% of the 
individual chromosome size,
respectively; 51\% of chromosome Y are unsequenced.

Our analysis on human DNA sequences is conducted using coarse-grained data.
Each original sequence was transformed into a spatial series
of GC content (GC\%) values. To this end, we evenly partition a
DNA sequence into $N$ non-overlapping windows of length $w$ bases, 
compute $\rho_i(w)=$GC\%$_i$ for each window $i$, to obtain a spatial
GC\% series: 
\begin{equation}
\label{rho}
\{ \rho_i \} \equiv  \{ \rho_i(w) \}  \equiv \{ \mbox{GC\%}_i \}
\hspace{0.1in}
\mbox{i= 1, 2, $\dots$, N }
\end{equation}
Table~1 lists the corresponding window sizes 
for each human chromosome.  Since different human chromosomes have 
different sizes, whereas the number of partitions ($N$) is the same, 
the window lengths vary.

\begin{table}[ht]
\caption{\label{tab:table1}
Average GC content ($\overline{ \mbox{GC}\%}$ or $\overline{ \rho }$), the window 
size ($w$) for partitions using $N=2^{17}$ non-overlapping windows for
twenty-four human chromosomes. Low-frequency scaling exponents 
$\alpha_1$ are estimated  from $S(f; s=3) \sim 1/f^{\alpha_1}$
in the range of $10^{-7} < f < 10^{-5}$ base$^{-1}$, and high-frequency
scaling exponents $\alpha_2$  are estimated in the range of
$10^{-5} < f < 2 \times 10^{-4}$ base$^{-1}$. The difference
between the two scaling exponents, $\Delta \alpha \equiv \alpha_2-\alpha_1$,
are listed in the fifth column. Low- and high-frequency exponents for $S(f)$
with substituted interspersed repeats are indicated by
$\alpha'_1$ and $\alpha'_2$, and their difference by
$\Delta \alpha' \equiv \alpha'_2-\alpha'_1 $.
}
\begin{ruledtabular}
\begin{tabular}{l|c|c|c|c|c|c|}
Chr & $\overline{ GC\%}$ & $w$ (kb) & $\alpha_1$ $\alpha_2$ & $\Delta \alpha$ 
 & $\alpha'_1$ $\alpha'_2$ & $\Delta \alpha'$ \\
\hline
1  & 41.7 &1.88 & 0.88 0.46 & 0.42 & 0.80 0.29 &0.51\\
2  & 40.2 &1.86 & 0.99 0.51&  0.48 & 0.96 0.30 &0.66\\
3  & 39.7 &1.52 & 0.95 0.43& 0.53 & 0.88 0.27 &0.61 \\
4  & 38.2 &1.46 & 0.87 0.34& 0.53 & 0.75 0.19 &0.57\\
5  & 39.5 &1.38 & 0.89 0.39& 0.51 & 0.88 0.23 &0.65 \\
6  & 39.6 &1.30 & 0.99 0.36& 0.63 & 0.86 0.24 &0.63\\
7  & 40.7 &1.21 & 0.97 0.46& 0.51 & 0.87 0.33 &0.55 \\
8  & 40.1 &1.12 & 0.97 0.42& 0.55 & 0.91 0.26 &0.66\\
9  & 41.3 &1.04 & 0.96 0.39& 0.57 & 0.90 0.28 &0.62 \\
10 & 41.6 &1.03 & 0.97 0.52& 0.46 & 0.95 0.34 &0.61 \\
11 & 41.6 &1.03 & 1.05 0.50& 0.55 & 0.97 0.35 &0.62 \\
12 & 40.8 &1.01 & 0.97 0.39& 0.59 & 0.89 0.28 &0.61 \\
13 & 38.5 & 0.86 & 0.83 0.33 & 0.50 & 0.73 0.24 &0.49 \\
14 & 40.9 &0.80  & 1.03 0.36 & 0.66 & 0.95 0.27 &0.68\\
15 & 42.2 &0.76 & 0.90 0.50 & 0.40 & 0.83 0.39 &0.44 \\
16 & 44.8 &0.69 & 0.91 0.51 & 0.40 &0.81 0.36 &0.45\\
17 & 45.5 &0.62 & 0.98 0.57 & 0.42 & 0.89 0.44 &0.46 \\
18 & 39.8 &0.58 & 1.12 0.40 & 0.72 & 1.12 0.28 &0.83 \\
19 & 48.4 &0.49  & 1.00 0.56 & 0.44 & 0.81 0.37 &0.45 \\
20 & 44.1 &0.49  & 0.87 0.51 & 0.36 & 0.83 0.30 &0.53 \\
21 & 40.9 &0.36 & 0.91 0.33 & 0.58 & 0.86 0.22 &0.64 \\
22 & 47.9 &0.38  & 0.90 0.62 & 0.28 & 0.86 0.40 &0.45 \\
X  & 39.4 &1.17  & 0.93 0.38 & 0.54 & 0.73 0.18 & 0.55 \\
Y  & 39.1 &0.38  & 0.83 0.38 & 0.45 & 0.70 0.21 & 0.49 \\
\end{tabular}
\end{ruledtabular}
\end{table}

Human DNA sequences contain a large fraction of interspersed repeats,
i.e., copies of an ancestral sequence fragment that possess a high
similarity between the duplicated and the ancestral sequence.  One can
detect interspersed repeats by using the program {\sf RepeatMasker}
\cite{repeatmasker}. ``Soft-masked'' annotations of interspersed repeats 
are taken from the DNA sequences of the UCSC human genome repository
({\sf http://genome.ucsc.edu/}), where repetitive (non-repetitive)
bases are annotated in small (capital) letters. Figure~\ref{fig-dirk1} 
shows the length distribution of the three sequences classes of uninterrupted
non-repetitive, interspersed repeat, and gap sequences.
Figure~\ref{fig-dirk2} shows the corresponding distribution of the
genome-wide GC\% for these three sequences classes.

To investigate the effect of interspersed repeats, we substitute 
them by random bases according to the chromosomal level of GC\%. 
Transformed, repeat-substituted DNA sequences of original human 
chromosomes are distinguished from original sequences.  On the 
coarse-grained level, it is equivalent to
the replacement in the $\{ \rho_i \}$ ($i=1, 2, \dots, N$) series
of any values calculated from the interspersed repeats by a random
value which is sampled from a Gaussian distribution; the
mean and variance of this Gaussian distribution is the
same as those of GC\% in the original sequence.  Another possibility
consists in substituting repetitive sequences by
by a constant value (e.g., the averaged GC\% value
of the original sequence).  This method introduces
additional correlations (and less variance) in the $\{ \rho_i \}$
series,  and is not adopted in this paper.
 
Three different, albeit functionally related, measures are
applied to the $\{ \rho_i \}$ series: the power spectrum
as a function of the frequency $S(f)$, the correlation function
$\Gamma(d)$ as a function of the distance $d$ between
windows, and variance $\sigma^2(w)$ of GC\%
series as a function of the window size $w$.

First, we conduct spectral analyses by calculating the power spectrum, 
the absolute squared-average of the Fourier transform, defined as:
\begin{equation}
S(f) \equiv \frac{1}{N} \left| \sum_{k=1}^{N} \rho_k
\cdot
e^{ -i 2 \pi k f/N} \right|^2.
\end{equation}
where $N$ is the total number of windows, and $f$ is measured
in units of cycle/window, which can be converted to units
of cycle/base by the window size (cf. Table~1). 

Coarse-graining ``hides'' base-base correlations at scales smaller
than $w$ bases.  The choice of $N = 2^{17}$ windows was made such 
that it is (i) sufficiently large to cover small-scale fluctuations, 
while (ii) at the same time sufficiently small so that the spectral analysis is
computationally feasible. As different chromosomes have difference
lengths, equal number of partitions leads to different window sizes $w$. 

The unsmoothed $S(f)$, or periodogram, contains $N/2$ independent
spectral components.  One can filter periodograms to obtain a
``smoothed'' spectrum $S(f;s)$, where $s$ is the span-size
parameter.  Since filtering with a relatively large $s$-value 
possibly distorts the shape of $S(f;s)$ at lower frequency components, 
different span-sizes are applied for different frequency ranges.

The second measure applied to the $\{ \rho_i \}$ series
is the correlation function, $\Gamma(d)$, which is computed
from two truncated series 
of $\{ \rho_i \}$, $ \rho' = \{ \rho_k \}$  ($k=1, 2, \dots, N-d$) and 
$ \rho'' = \{ \rho_k \}$  ($k=d+1, d+2, \dots,  N$):
\begin{equation}
\label{gamma}
\Gamma(d) \equiv \frac{ \rm Cov(\rho', \rho'')}
{ \sqrt{ \rm Var(\rho')} \sqrt{ \rm Var(\rho'')}}
\end{equation}
where $\mbox{Cov}( \rho', \rho'')=
\langle \rho' \rho''\rangle - \langle \rho'\rangle \langle \rho''\rangle $ 
and $\mbox{Var}( \rho') = \langle \rho'^2 \rangle -  \langle  \rho' \rangle^2$ 
(or $\mbox{Var}( \rho'') = \langle \rho''^2 \rangle -  \langle  \rho'' \rangle^2$) 
are the covariance and variance.
Note that the $\Gamma(d)$ defined in Eq.(\ref{gamma})
is slightly different from that defined using
a periodic boundary condition.

The third and final measure applied to the 
$\{ \rho_i \}$ series is the variance $\sigma^2(w)$:
\begin{equation}
\sigma^2(w) \equiv \langle \rho(w)^2 \rangle -
\langle \rho(w) \rangle^2
\end{equation}
as a function of the window size $w$.
The power spectrum, the correlation function, and the window-size-dependent
variance are interrelated quantities \cite{clay3}:
\begin{equation}
  \sigma^2(w) \sim \frac{ \Gamma(0) }{ w } \cdot 
\bigg\{ 1 + \frac{2}{w}\sum_{d=1}^{w-1} (w-d) \Gamma(d) \bigg\}.
\end{equation}
If $S(f) \sim 1/f^\alpha$, $\Gamma(d) \sim 1/d^\gamma$,
$\sigma^2(w) \sim 1/w^\beta$ are power-law functions, 
then their scaling exponents are related  
by $\alpha = 1-  \gamma$ and  $\gamma=\beta$ \cite{clay3}.

The calculation of $S(f)$ and $\Gamma(d)$ was carried out by the 
statistical package {\sl S-PLUS} (Version 3.4, MathSoft, Inc.), and the
type of filter implemented for $S(f)$ is the Daniell-filter \cite{daniell}.

\SEC{$1/f$ noise  is a universal feature of human DNA sequences}

In this section, we use the power spectrum $S(f)$ to study
GC\% of human genome sequences, with
the goals of testing the universality of $1/f$ noise, quantifying
different decay ranges for $S(f) \sim 1/f^\alpha$, and comparing
$S(f)$ across DNA sequences of different human chromosomes.

\begin{figure*}
\centerline{\psfig{figure=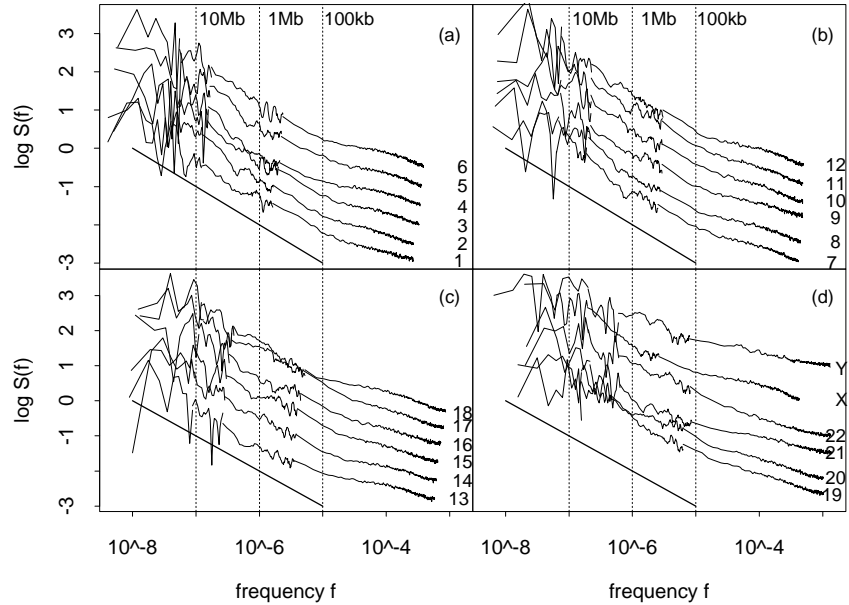,width=80mm,angle=-90}}
\caption{\label{fig3:s(f)} 
Double-logarithmic representation of 
power spectra $S(f)$ of GC\% of all twenty-four human 
chromosomes. Each plot shows $S(f)$ of six chromosomes
(shifted on the $y$-axis for clearer representation):
chromosomes (a) 1--6; (b) 7--12; (c) 13--18; (d) 19--22, X, and Y. 
The $x$-axis (in logarithmic scale) is converted from cycle/window 
to cycle/base by using the window sizes listed in Table~1.
$S(f)$ is filtered at different levels for different frequency
ranges: $S(f; s=1)$ for the first ten spectral components,
$S(f; s=3)$ for the components 11--30,
$S(f; s=31)$ for the components 31--400,
and $S(f; s=501)$  for the components 400--65536 (=$2^{16}$).
}
\end{figure*}

Figure~\ref{fig3:s(f)} shows for $N=2^{17}$ GC\% values the power 
spectra $S(f)$ across all human chromosomes. We find that $S(f)$ 
exhibits no clear plateau at 
low frequency ($< 10^{-6}$ cycle/base) and increases steadily
with decreasing frequency. The decay can be mathematically 
approximated by a power-law of the form $S(f) \sim 1/f^\alpha$ 
with $\alpha \approx 1$.  Table~1 lists for the frequency range
$f=$ 10~Mb$^{-1}$--100~kb$^{-1}$ the estimated scaling exponent
$\alpha_1$ for all chromosomes, using a best-fit regression
of $\log_{10} S(f; s=3) = a + \alpha_1 \log_{10}(f)$. We find that
$\alpha_1$ is typically close to $\alpha_1 \approx 1$ with 
practically little variation across chromosomes.

\begin{figure}
\centerline{\psfig{figure=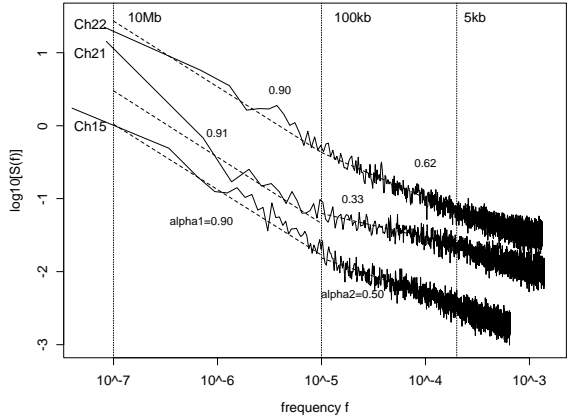,width=55mm,angle=-90}}
\caption{\label{fig4:ex} 
Cross-over from $S(f) \sim 1/f^{\alpha_1}$ to $S(f) \sim 1/f^{\alpha_2}$ 
illustrated for human chromosomes 15, 21, and 22
(smoothed with the span size of 31, and shown in double-logarithmic scale).
The scaling exponents $\alpha_1$ and $\alpha_2$ are shown  for 
the frequency ranges 10~Mb$^{-1}$--100~kb$^{-1}$ and 100~kb$^{-1}$--5k$^{-1}$.
}
\end{figure}

A closer inspection of Fig.~\ref{fig3:s(f)} shows that the 
majority of $1/f$ spectra undergo a cross-over from $\alpha_1 \approx 1$ 
to $\alpha_2 < 1 $ at high frequency. The deviation from $\alpha_1 \approx 1$ 
starts about 30--100~kb and continues at smaller distances.
Figure~\ref{fig4:ex} illustrates this feature for $S(f; s=31)$ 
of the DNA sequences of Chr15, Chr21, and Chr22 in more detail.
We find that chromosomes 15 and 21 exhibit clear cross-overs 
at about 100~kb, while chromosome 22 exhibits no apparent break-point.
Table~1 contains for the frequency range of 
$f=$ 100~kb$^{-1}$--5~kb$^{-1}$ the corresponding scaling
exponents $\alpha_2$, obtained from the 
regression $\log_{10} S(f; s=3) = a + \alpha_2 \log_{10}(f)$. 
We find a pronounced difference in absolute values between 
$\alpha_1 \approx 1$ and $\alpha_2 < 1$, indicating a transition
from the universal $1/f^{\alpha_1}$ ($\alpha_1 \approx 1 $) 
spectrum at low frequency to a 
more flattened $1/f^{\alpha_2}$ ($\alpha_2 < 1$) spectrum
at higher frequency.

\begin{figure}
\centerline{\psfig{figure=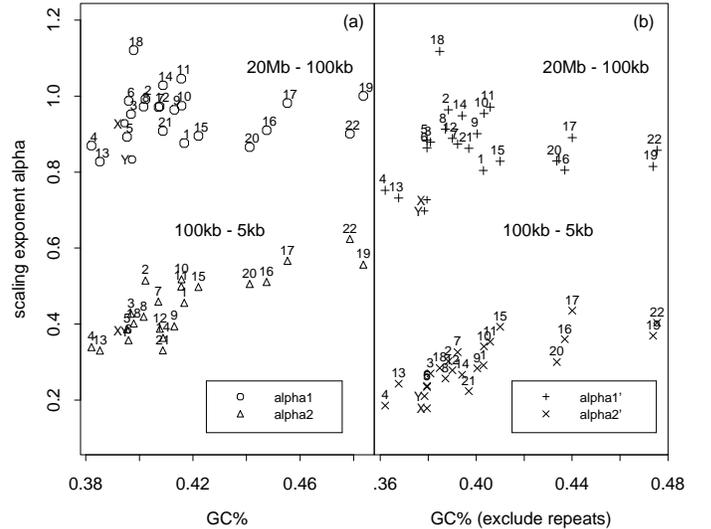,width=70mm,angle=-90}}
\caption{\label{fig5:alpha} 
(a) Scaling exponents $\alpha_1$ and $\alpha_2$
for fitting the power spectrum $S(f) \sim 1/f^{\alpha_i}$ 
($i=1,2$) at the frequency range of 10~Mb$^{-1}$--100~kb$^{-1}$,
and 100~kb$^{-1}$--5~kb$^{-1}$, respectively, versus the
chromosome-specific GC content of all 24 human chromosomes.
(b) Scaling exponents $\alpha'_1$ and $\alpha'_2$
for $S(f)$ with substituted interspersed repeats.
}
\end{figure}

Figure~\ref{fig5:alpha}(a) shows for all human chromosomes $\alpha_1$ 
and $\alpha_2$ as a function of chromosome-specific GC\%. The
majority of human chromosomes have a specific GC content ranging between
38--43\%, whereas chromosomes 16, 17, 19, 20, and 22 have higher GC\%
up to 49\%.  While the low-frequency scaling exponent $\alpha_1$ 
remains approximately independent of GC\%, Fig.~\ref{fig5:alpha}(a)
shows that $\alpha_2$ increases with increasing GC\% and
gives rise to a positive correlation between $\alpha_2$ and GC\%.

The three chromosomes illustrated in Fig.~\ref{fig4:ex} exhibit
different degrees of transition from the $1/f^{\alpha_1}$
($\alpha_1 \approx 1$) to the flattened
$1/f^{\alpha_2}$ ($\alpha_2 <1$) spectrum, with chromosome 21 (22)
undergoing the sharpest (smoothest) transition. This
observation can be further quantitized by the change in
scaling exponents $\alpha_1$ and $\alpha_2$. Table~1 lists for
all chromosomes $\Delta \alpha = \alpha_2 - \alpha_1$. 
Chromosome 22 is distinct from all other human chromosomes
as the most scale-invariant one (same or similar scaling 
exponent at different length scales).
The same observation that human chromosome 22 was perhaps different 
from the remaining human chromosomes was made using limited 
sequence data in \cite{isochore-clay,pedro}. 

\SEC{Interspersed repeats are not responsible for 
$1/f$ spectrum}

About 45\% of human genomic DNA sequences are interspersed 
repeats \cite{lander}. Interspersed repeats consist of copies of the same
sequence segment that are inserted in the human genome, possess a high
similarity between the duplicated and ancestral sequence, and have
been implicated in a variety of biological functions, including genome
organization, human chromosome segregation, or regulation of gene
expression \cite{repeats-bio}. Large copy numbers increase 
the sequence redundancy and it has been shown, e.g., that
about 10\% interspersed Alu repeats significantly increase
base-base correlations in the range up to 300~bases
\cite{dirk}.

\begin{figure}
\centerline{\psfig{figure=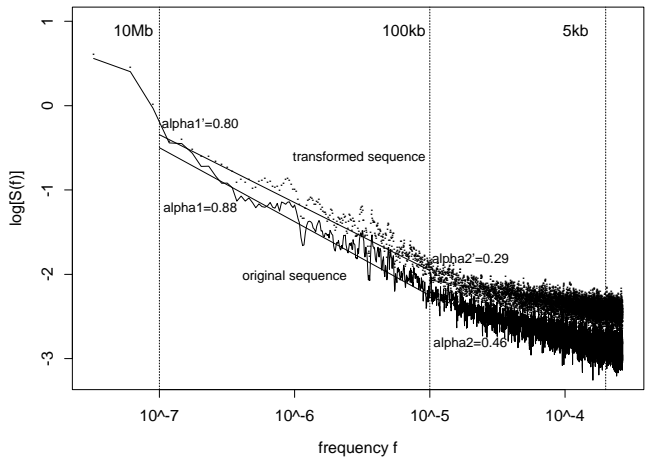,width=60mm,angle=-90}}
\caption{\label{fig6:rep} 
Power spectra $S(f)$ of GC\% for the original and the
transformed (interspersed repeats substituted)  DNA 
sequence of human chromosome 1.  The scaling exponent for 
low-frequency (10~Mb--100~kb) and high-frequency
(100~kb--5~kb) ranges are obtained by a best-fit regression
of $\log_{10} S(f)$ over $\log_{10} f$.
}
\end{figure}

Figure\ref{fig6:rep} shows the power spectrum $S(f)$ for 
the original human chromosome 1 and for the transformed sequence in 
which interspersed repeats are substituted.  We find in 
the low-frequency range of $10^{-7} < f < 10^{-5}$ cycle/base 
that $S(f)$ decays in the original sequence
with $\alpha_1 \approx 0.88$ and in the transformed sequence with
$\alpha^\prime_1 \approx 0.80$, indicating only marginal differences in
the decay properties of $S(f)$ due to repetitive sequences.  In
contrast, in the high frequency range of $10^{-5} < f < 2 \times 10^{-4}$ 
we find $\alpha_2 \approx 0.46$ and $\alpha^\prime_1 \approx 0.29$,
and thus interspersed repeats contributes to the decay properties 
of $S(f)$ for high-frequency components by flattening the power spectrum.

The scaling exponents $\alpha^\prime_1$ and $\alpha^\prime_2$ for 
repeat-substituted DNA sequences of all 24
human chromosomes are shown in Table~1.  The difference
between low- and high-frequency ranges for DNA sequences of original
chromosomes, $\Delta \alpha= \alpha_2 - \alpha_1$, is smaller 
than the difference between low- and high-frequency ranges for 
transformed sequences, $\Delta \alpha' = \alpha^\prime_2-\alpha^\prime_1$.  
When we compare $\alpha_1$ and $\alpha^\prime_1$, as well as $\alpha_2$ and
$\alpha^\prime_2$, we find that the magnitude of $\alpha^\prime_1$
($\alpha^\prime_2$) is always smaller than that of $\alpha_1$
($\alpha_2$), which means a flattened spectrum
due to the substitution of interspersed repeats.  
The average change of low-frequency
scaling exponents, $ \alpha_1- \alpha'_1$, is about 0.07, 
whereas the average change of high-frequency scaling
exponents, $ \alpha_2- \alpha'_2$, is about 0.14. This
confirms that the universal presence of $1/f$ spectrum
at low frequency is not caused by interspersed
repeats,  but that interspersed repeats affect $S(f)$
predominantly  at high frequencies.  A similar conclusion 
that the decay rate of base-base correlations in DNA sequences of
human chromosomes 20, Chr21, and Chr22 is not markedly affected
by the substitution of interspersed repeats was reached in \cite{dirk}.

We note that the extent of deviation,
$|\alpha'-\alpha|$, depends on how the replacement of 
interspersed repeats  is conducted. Possible substitutions
of interspersed repeats include the substitution by
a constant value or a randomly sampled value.
In general, the substitution of GC\% values
calculated from the repetitive sequences by random values enhances 
the deviation and flattens the spectrum $S(f)$ more than the
substitution by a constant value (e.g., average GC\%).

\SEC{Resistance to variance reduction at larger window sizes}

In this section, we study the decay properties of the
variance ($\sigma^2$) of spatial GC\% series  as 
a function of difference window sizes $w$,  and
we compare the scaling of $\sigma^2$ with the
scaling of the power spectrum $S(f)$.

Early experimental measurement of the GC\% distribution by
using cesium chloride (CsCl) profile \cite{cscl} showed for mouse 
{\em Mus musculus} genomic DNA sequences that the 
variance of GC\% values does not markedly decreases with the DNA segment size
 \cite{macaya}. This experimental observation is directly related 
to the presence of 1/f spectra in DNA sequences 
\cite{isochore-clay,li-nova}. If the variance of the spatial 
GC\% series calculated at the window size $w$ is $\sigma^2(w)$, 
then a scaling of 
$\sigma^2(w) \sim 1/w^\beta$ implies a corresponding 
scaling in the power spectrum $S(f) \sim 1/f^{1-\beta}$ 
\cite{isochore-clay,beran}. If GC\% is obtained from 
$w$ uncorrelated bases, it follows a binomial distribution.
Consequently, $\sigma^2(w) \sim \langle\rho \rangle 
(1- \langle \rho \rangle) /w \sim 1/w$ with 
$\beta=1$. The corresponding scaling exponent of
the power spectrum is $\alpha=1-\beta=0$, and thus the
$S(f) \sim \mbox{cons.}$ is equivalent to the white noise.

\begin{figure}
\centerline{\psfig{figure=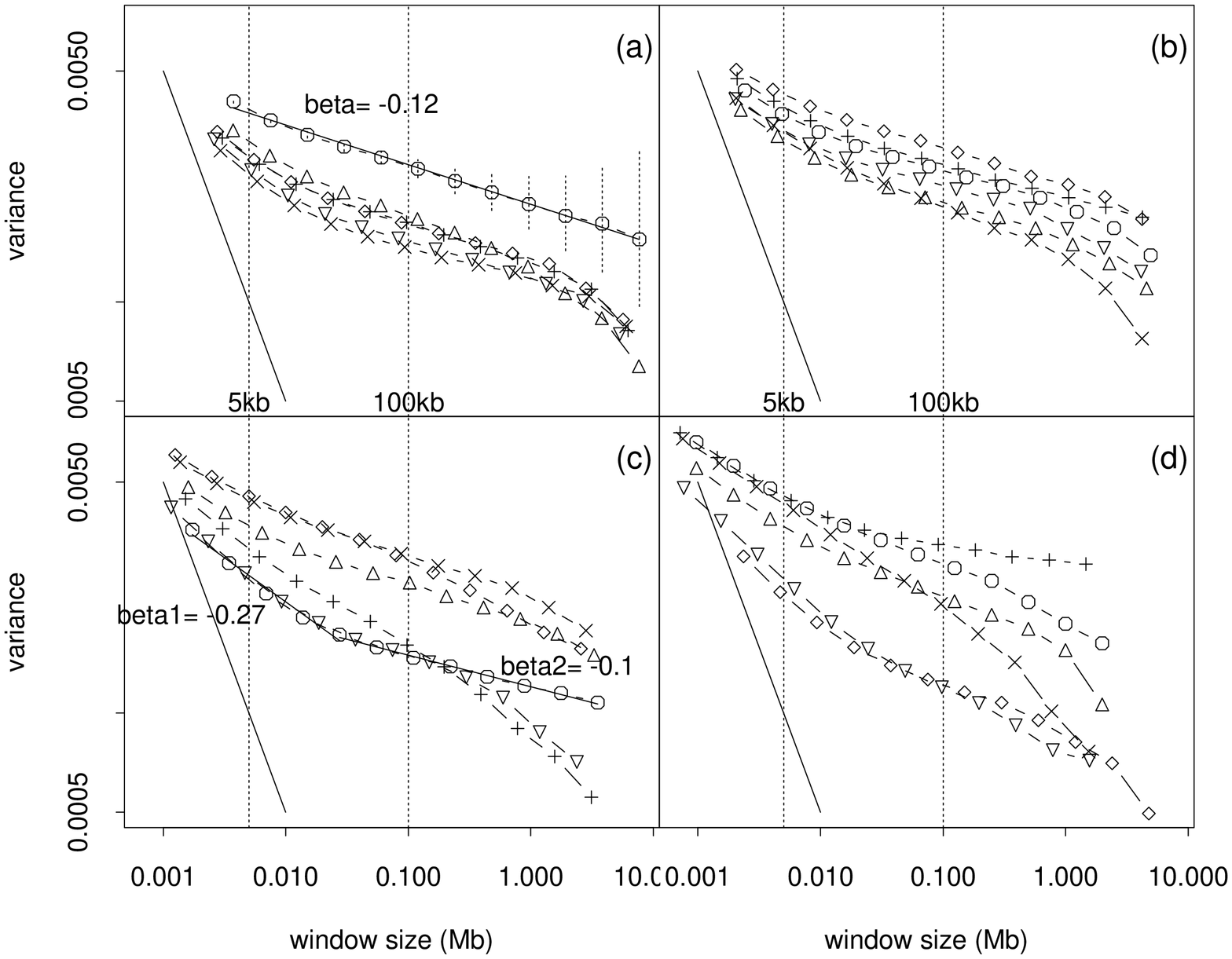,width=65mm,angle=-90}}
\caption{\label{fig7:var} 
Double-logarithmic representation of the variance $\sigma^2(w)$ 
of the spatial GC\% series for all human chromosomes (Chr)
as a function of the window size $w$: 
(a) $\bigcirc$ Chr1,
$\triangle$ Chr2,
$+$ Chr3, $\times$ Chr4, $\diamondsuit$ Chr5, $\bigtriangledown$ Chr6;
(b) $\bigcirc$ Chr7, $\triangle$ Chr8, $+$ Chr9, $\times$ Chr10,
$\diamondsuit$ Chr11, $\bigtriangledown$ Chr12;
(c)  $\bigcirc$ Chr13, $\triangle$ Chr14, $+$ Chr15,
$\times$ Chr16, $\diamondsuit$ Chr17, $\bigtriangledown$ Chr18;
(d)  $\bigcirc$ Chr19, $\triangle$ Chr20, $+$ Chr21, $\times$ Chr22,
$\diamondsuit$ ChX, $\bigtriangledown$ ChrY.
Straight lines indicate $\sigma^2(w) \sim 1/w$ (corresponding
to white noise).  One regression line for Chr1 ($\beta \approx $0.12)
and a piece-wise regression for Chr13 ($\beta \approx $0.27 and 
$\beta \approx $0.10) are drawn. The 95\% confidence
interval for the $\sigma^2(w)$ estimation of Chr1 at each point
of $w$ is marked by a vertical dashed line.
}
\end{figure}

Figure~\ref{fig7:var} shows $\sigma^2(w)$ as a function of window size $w$ 
for all human chromosomes. In a double-logarithmic representation,
we find that $\log (\sigma^2(w)) $ decays approximately
linearly with $\log( w) $. A decay according to  
$\sigma^2(w) \sim 1/w^\beta$ with $\beta=1$ leads to 
white noise. This situation is indicated in Fig.~\ref{fig7:var}
by the straight line. An inspection of Fig.~\ref{fig7:var}
shows, however, that the variance decays at a much slower 
rate than what would be for white noise. The variance of
the DNA sequence of human chromosome 1, e.g., 
gives rise to $\beta \approx 0.12$, and the
corresponding scaling exponent $\alpha_1 \approx
1- \beta =0.88$  is indeed close to the estimated
exponent listed in Table~1. The scaling of the variance
with the exponent $\beta  << 1 $ is in accord with
the low-frequency $1/f$ noise.

The scaling of $\sigma^2(w)$ shown in Fig.~\ref{fig7:var} 
differs from one human chromosome to another. For instance,
in the range of $w=$ 1~kb--5~Mb, for example, human chromosome 
13 exhibit a clear transition from $\beta_2 \approx 0.27$ ($w < $ 50~kb) 
to $\beta_1 \approx 0.10$ ($w > $ 50~kb), corresponding to 
$S(f) \sim 1/f^{0.63}$ and $S(f) \sim 1/f^{0.9}$,
respectively, at high- and low-frequency ranges.  
Other human chromosomes, although generally exhibiting a power-law 
scaling form of $\sigma^2(w)$, show deviations from 
$\sigma^2(l) \sim 1/l^{\beta}$ line for the largest
window sizes tested.

The investigation of $\sigma^2(w)$ as a function of
different window sizes $w$ requires careful examination
\cite{audit,pedro04}.  First, since we partition each
human chromosome in $2^k$ ($k=$17, 16, $\dots$) 
windows, the variance of GC\% series $\{ \rho_i \} $ 
could be accidentally large when windows reside on
the isochore borders, and small by chance if they
start/end within an isochore. 

Second, when the number of windows is small (e.g. the last 
point of $\sigma^2(w)$ for each chromosome in Fig.~\ref{fig7:var} 
is calculated with the largest window size that
gives rise to  32 windows), the standard error of the 
sample variance is large. The 95\% confidence interval for
$\sigma^2(w)$ of Chr1 is shown in Fig.~\ref{fig7:var}(a), 
using the interval:
[$ (w-1) \sigma^2/t_{0.025}, (w-1) \sigma^2 /t_{0.975}$],
where $t_x$ is defined by $\int_{-\infty}^{t_x} \chi^2(\rm{df}=w-1) dt= x$
(where $\chi^2(\rm{df})$ is the chi-square distribution with $\rm{df}$
degrees of freedom) \cite{snedecor}.  Figure~\ref{fig7:var}(a)
shows that for fewer windows (and larger window sizes), 
the 95\% confidence interval of $\sigma^2(w)$ could be large
such that the estimated value of $\beta$ may change from sample to sample.

Finally, the relationship between scaling exponents 
$\alpha+\beta = 1$ \cite{beran,isochore-clay},
is based on the assumption that both $S(f)$ and $\sigma^2(w)$ are 
theoretical power-law functions. If $S(f)$ is a piece-wise
power-law function, as in the case of GC\% fluctuation of
human chromosomes, a correction term to the relationship
$\alpha +\beta=1$ is expected.
 
\begin{figure}
\centerline{\psfig{figure=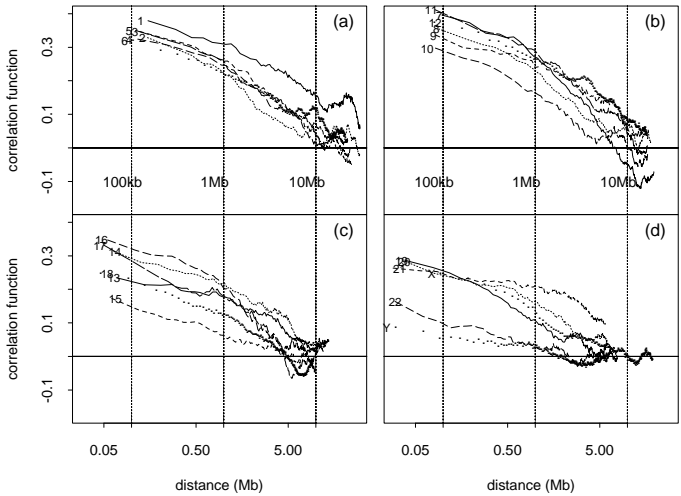,width=65mm,angle=-90}}
\caption{\label{fig8:corr} 
Correlation function $\Gamma(d)$ for 24 human chromosomes (Chr)
as a function of the window distance $d$ (converted to bases 
by the window size listed in Table~1).  The distance is
represented on a logarithmic scale. (a) Chr1--6; (b) Chr7--12; 
(c) Chr13--18; and (d) Chr19--22, ChrX, and ChrY.
}
\end{figure}

\begin{figure}
\centerline{\psfig{figure=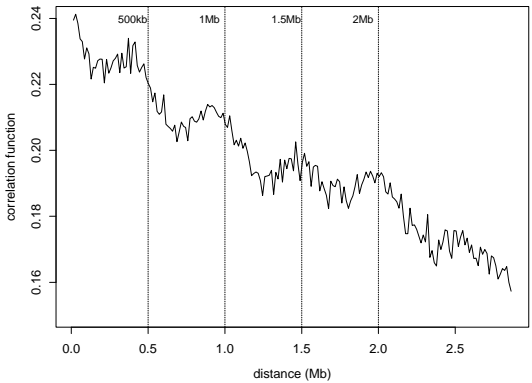,width=50mm,angle=-90}}
\caption{\label{fig9:ch21} 
Correlation function $\Gamma(d)$ for human chromosome 21
as a function of the window distance $d$  (converted to bases 
by the window size given in Table~1). The oscillation in $\Gamma(d)$
is highlighted by vertical lines, indicating the
distances of $d=$500~kb, 1~Mb, 1.5~Mb, and 2~Mb.
}
\end{figure}

\SEC{Chromosome-specific correlation structures}

Apparently, $1/f$ noise in music and speech signals \cite{voss-music}
does not prevent music and speech from sounding differently.
Similarly, universal $1/f^\alpha$ spectra in GC\% fluctuations
across human chromosomes do not imply that all chromosomes
exhibit the same detailed correlation structure. The generic trend
of $S(f)$ spectra to increase at low frequency may ``co-exist"
with small peaks at higher frequency.  Such chromosome-specific 
characteristic length scales can be more intuitively examined
by correlation functions. In this section, we investigate
the correlation function $\Gamma(d)$
of coarse-grained DNA sequences of human chromosomes with the
aim of further examining chromosome-specific structures,
such as characteristic length scales and oscillation
detected by $\Gamma(d)$.

Figure~\ref{fig8:corr} shows for all human chromosomes
the $\Gamma(d)$'s of GC\% series
$\{ \rho_i \}$ calculated for the window sizes given in Table~1,
of all human chromosomes.  For each chromosome, the minimum 
(maximum) distance is 80 (16,000) windows.
Since each chromosome is partitioned into $2^{17}$ windows, 
the maximum distance $d$ at which the correlation is examined 
is about $16,000/2^{17} \approx 12\%$ of the total sequence length.

An inspection of Fig.~\ref{fig8:corr} shows that the magnitude
of correlation at the distance of $d=1~{\rm Mb}$ is clearly above
the noise level. With the exceptions of Chr15, Chr22, and ChrY, 
the correlation function $\Gamma(d) > 0.1$ at $d=1~{\rm Mb}$
for all other chromosomes. The low correlation in ChrY is 
due to the fact that about half of the bases are unsequenced, 
and the substitution of gaps by random values lowers the correlation. 
At even longer distances such as $d=$10~Mb, correlations 
$\Gamma(d=10$~Mb) for chromosomes 1 and 6 are still
above the 0.1 level. 

Given different windows ($w$) due to different chromosome 
sizes and provided that the covariance of GC\% is approximately 
independent of $w$, a scaling of the variance according to 
$1/w^\beta$ implies that the  correlation function
$\Gamma(d)$ in Eq.(\ref{gamma}) increases with the window
size as $\sim w^\beta$. Test calculations of covariance 
for $2^{15}$ and $2^{17}$ windows show that the covariance 
differs by less than 1\% (and hence is fairly independent in 
this range of window sizes).
Consequently, for a detailed comparison of correlation 
functions calculated for different chromosomes one has to 
take into account different windows sizes.

Any deviation from the monotonic decrease of $\Gamma(d)$ 
might be indicative of correlations at characteristic
length scales (visible as ``bumps"). For example, 
Fig.~\ref{fig8:corr} shows for chromosome 1 such a bump at 
$d \approx$ 21--23~Mb. Bumps or sharper peaks in other chromosomes include 
$d \approx$ 9.3~Mb (Chr2), 7.2~Mb (Chr10), 3.2--3.8~Mb 
(Chr12), and 2.4--3.1~Mb (Chr19).  One  plausible
explanation is that for chromosomes 2, 10, 12, and 19 one or few
alterations of GC-rich/low isochores \cite{isochore} with these 
length scales enhance the correlation.

Chromosome 21 stands out among all human chromosomes for
having a comparatively higher correlations at distances of
several Mb (despite having a smaller $w^\beta$ factor than 
other chromosomes due to a smaller window size).  A detailed 
inspection of Fig.~\ref{fig9:ch21} uncovers an oscillation of 
$\Gamma(d)$ of  about 500~kb, ranging from $d=$500~kb to $d=$2~Mb, which has
not been reported before. It can be further shown that this 
oscillation is not due to the substitution of interspersed repeats 
\cite{li-holste-21},  and it is localized to about one-eighth 
of the right distal end of chromosome 21  \cite{li-holste-21}.

\SEC{Discussions}

We study correlation structures and spectral components in 
the set of human chromosomes, using power spectra, 
coarse-grained correlation functions, and the variance 
of different window sizes. All three measures are 
interrelated and highlight compositional structures 
at different feature levels.  Our results firmly establish 
the presence of long-ranging correlations and 
$1/f^\alpha$ spectra in the DNA sequences of the set 
of twenty-four  human chromosomes. 

Using updated and completed human sequence data, we find the 
presence of 1/f noise in the DNA sequences of 
all human chromosomes. We further find that, with the exception 
of chromosome 22, all chromosomes exhibit a cross over 
from $1/f^{\alpha_1}$ at low-frequency to $1/f^{\alpha_2}$ 
scaling at high-frequency ($\alpha_1 > \alpha_2$).
The result of two scaling ranges at low- and high-frequency 
are in accord with previous findings, obtained from 
sequence data of lower quality, and it refines break-point 
regions for each individual chromosome.

We also examined the effect of about 45\% interspersed 
repeats in the human genome. Using a procedure in which 
masks and subsequently substitutes interspersed repeats 
with random GC\% values, we find that interspersed 
repeats (i) only marginally affect the scaling exponent 
$\alpha_1$ in the low-frequency range, but (ii) lower 
$\alpha_2$ in the high-frequency range (cf.~Fig.\ref{fig5:alpha}(b)).
This supports the general understanding that interspersed repeats 
only contribute to short-ranging (high-frequency) correlations \cite{dirk}.

We have shown elsewhere that $1/f^\alpha$
spectra of GC\% fluctuation are also universally present
in the mouse {\sl Mus musculus}  genomic DNA sequences
\cite{1fmouse}. It is known that human and mouse genomes 
are separated by approximately 65--75 million years of evolution. Besides 
the similarity (or homology ) between
these two genomes on a local scale, there is in fact a
large amount of reshuffling of the chromosome segments
at a global scale when two current-day copies of
the two genomes are compared side-by-side \cite{pevzner}. Since reshuffling
of a sequence at global scales could potentially destroy
long-range correlations, it is still to be resolved
under what conditions a reshuffling of the 
human genome into the mouse genome, or vice versa,
conserves 1/f noise.

One possible explanation of why $1/f^\alpha$ spectra appear
in both the human and the mouse genomes is that such long-range
patterns were probably generated from ancestral DNA
sequences by sequence evolutionary mechanisms. One
sequence evolution model, termed expansion-modification
(EM) model, is known to generate $1/f^\alpha$ spectra \cite{wli-em}.
The EM model incorporates duplications and mutations. 
Since the duplication process is an essential element in 
evolutionary genomics \cite{ohno}, whose role is perhaps as important as Darwin's 
natural selection \cite{meyer}, even a yet unsophisticated 
incorporation of duplications in the EM model may capture the 
essence of the evolutionary origin of long-range correlations
in DNA sequences. In the EM model, only the duplication of
segments with the same length scale is included, whereas
in reality segments with a broad range of length scales
are duplicated \cite{lander}.

One frequently posed question concerns the ``biological meaning"
of $1/f^\alpha$ spectra or long-range correlations in DNA 
sequences. In order to address this question, one may ask
a couple of related questions beforehand. 
Does the compositional GC\% have any biological effects?
What biological functions of the DNA molecule are of relevance?
From the {\sl functional genomics} perspective, interesting
biological processes related to DNA molecules include transcription, 
replication, and recombination, and  their potential connection 
to GC\% has been reviewed in \cite{bernardi,bernardi-book,li-nova}.
Generally speaking, GC\% has a statistical association with 
all three processes, though the cause-and-effect role has
not yet been firmly established. Recent studies show that broadly
expressed ``housekeeping genes" tend to be located in GC-rich
regions \cite{housekeeping}. To understand the genome-wide organization 
of biological units that play a role in those processes
(e.g., genes, origins and timing of replication, or recombination hotspots), 
at times it is more feasible to directly study the
spatial distribution of functional units instead of using 
the GC\% as a surrogate. 

From the {\sl biophysics and cellular biology}  
perspective, GC\% is linked with  bands from chromosome-staining 
\cite{gojobori}, and in addition, possibly with the 
matrix/scaffold attachment/associated regions located at
the end of DNA loops \cite{attachment}.  It has also been
suggested that GC-rich chromosomes (or regions)
tend to be located in the interior of the nuclear 
during interphase and are more ``open" in their 
tertiary structure, whereas GC-poor segments are
more likely to be close to the surface of the
nuclear and more condensed \cite{interphase}.

Further exploration of the relationship between GC\% fluctuations,
as well as its large-scale patterns, and the above
biological processes is beyond the scope of this
paper. An attempt for bacterial genomes has been
made to relate the scale-invariance feature in sequence 
statistics to the genome organization of transcription 
activities \cite{audit}. It is clear that more 
integrated computational and experimental analyses 
need be carried out along similar lines 
before one can give universal $1/f$ spectra in 
DNA sequences a satisfactory biological explanation.

\SEC*{Acknowledgements}

We thank S. Guharay for participating the early stage 
of this project, and O. Clay, J.L. Oliver, A. Fukushima for
valuable discussions.

\end{document}